\documentclass[aps,prb,twocolumn,amsfonts,superscriptaddress,amssymb,amsmath]{revtex4-1}

\usepackage{graphicx}
\usepackage{siunitx}
\usepackage{subcaption}
\usepackage{enumitem}
\usepackage{bbold}
\usepackage{tikz}
\usepackage{color}
\newcommand{\eij}{$\epsilon_{ij}$}

\begin{document}


\title{Ultrafast optical excitation of coherent magnons in antiferromagnetic NiO} 


\author{Christian Tzschaschel}
\email{christian.tzschaschel@mat.ethz.ch}
\affiliation{Department of Materials, ETH Zurich, 8093 Zurich, Switzerland}

\author{Kensuke Otani}
\affiliation{Institute of Industrial Science, The University of Tokyo, Tokyo 153-5805, Japan}

\author{Ryugo Iida}
\affiliation{Institute of Industrial Science, The University of Tokyo, Tokyo 153-5805, Japan}

\author{Tsutomu Shimura}
\affiliation{Institute of Industrial Science, The University of Tokyo, Tokyo 153-5805, Japan}

\author{Hiroaki Ueda}
\affiliation{Department of Chemistry, Kyoto University, Kyoto 606-8502, Japan}

\author{Stefan G\"unther}
\affiliation{Department of Materials, ETH Zurich, 8093 Zurich, Switzerland}

\author{Manfred Fiebig}
\affiliation{Department of Materials, ETH Zurich, 8093 Zurich, Switzerland}

\author{Takuya Satoh}
\affiliation{Department of Materials, ETH Zurich, 8093 Zurich, Switzerland}
\affiliation{Institute of Industrial Science, The University of Tokyo, Tokyo 153-5805, Japan}
\affiliation{Department of Physics, Kyushu University, Fukuoka 819-0395, Japan}


\date{\today}

\begin{abstract}

In experiment and theory, we resolve the mechanism of ultrafast optical magnon excitation in antiferromagnetic NiO. We employ time-resolved optical two-color pump-probe measurements to study the coherent non-thermal spin dynamics. Optical pumping and probing with linearly and circularly polarized light along the optic axis of the NiO crystal scrutinizes the mechanism behind the ultrafast optical magnon excitation. A phenomenological symmetry-based theory links these experimental results to expressions for the optically induced magnetization via the inverse Faraday effect and the inverse Cotton-Mouton effect. We obtain striking agreement between experiment and theory that, furthermore, allows us to extract information about the spin domain distribution. We also find that in NiO the energy transfer into the magnon mode via the inverse Cotton-Mouton effect is about three orders of magnitude more efficient than via the inverse Faraday effect.

\end{abstract}


\maketitle


\section{Introduction}

Antiferromagnetism is rapidly gaining importance as a crucial ingredient of spintronics applications.\cite{Jungwirth16, Gomonay14} Because of the absence of a net magnetization in the ground state, it is robust against externally applied fields and the formation of domains is not obstructed by magnetic stray fields. Accordingly, the technologies envisaged are mainly based on the application of spin currents instead of magnetic fields.\cite{Zelezny14,Wang14,Moriyama15,Prakash16,Lin16,Khymyn16,Rezede16} In addition, the intimate coupling of the sublattice magnetizations in antiferromagnets in combination with a strong exchange interaction between neighboring spins implies magnetization-dynamical timescales, which are typically orders of magnitude faster than those of ferro- or ferrimagnetic materials.\cite{Satoh07} Naturally, ultrashort laser pulses come to mind when accessing the dynamical properties of the antiferromagnetic order. In contrast to thermal approaches, which are based on local heating of the electronic and magnetic systems,\cite{Manz16} non-thermal excitations would provide a quasi-instantaneous access to the antiferromagnetic spin system via spin-orbit coupling. Thus, they can fully exploit the faster timescales inherent to antiferromagnets. The two most prominent non-thermal magneto-optical effects are the inverse Faraday effect (IFE)\cite{Kimel05} and the inverse Cotton-Mouton effect (ICME).\cite{Kalashnikova07} Microscopically, they represent impulsive stimulated Raman scattering processes, where the IFE is described by an antisymmetric tensor and the ICME by a symmetric tensor.\cite{Shen66,Pershan66,Kalashnikova15} Consequently, the magneto-optical coupling effectively exerts a torque onto the spin system.

The IFE and ICME have been applied to a variety of material systems, \cite{Stanciu07,Kalashnikova08,Kimel09,Kirilyuk10,Satoh12,Ivanov14,Bossini17} but a clean discrimination in experiment and theory between the two effects for a pure antiferromagnet is still due. A particularly obvious candidate for such an analysis is antiferromagnetic NiO because of its high ordering temperature, its simple crystallographic structure, and its well-researched physical properties.\cite{Newman59,RothSlack60,Kondoh60,Roth60,KondohTakeda64,Hutchings72, Grimsditch98,Fiebig01,Milano04,Sanger06} In addition, it may be an excellent candidate for a clear and insightful experimental and theoretical discrimination between IFE and ICME because it has been speculated that in NiO the symmetric part is significantly larger than the antisymmetric part of the Raman scattering tensor.\cite{Grimsditch98} Consequently, the ICME would be more pronounced than the IFE, even though the ICME is a second-order effect in the magnetic order parameter. Unfortunately, the pronounced magnetic birefringence of NiO\cite{Roth60} leads to an inseparable mixture of the polarization-dependent Raman contributions. Hence, the spin oscillations observed in NiO are to date generally induced by an inseparable mixture of IFE and ICME. Consequently, the mechanism behind the non-thermal excitation of coherent magnons in NiO has not been identified, let alone quantified.\cite{SatohPRL10,Nishitani10,Kanda11,Higuchi11,Nishitani12,Takahara12}

In this Report, we present a comprehensive experimental and theoretical analysis of IFE and ICME in antiferromagnetic NiO. We separate the two effects in a non-thermal polarization-dependent two-color pump-probe measurement. The birefringence resulting from the optical anisotropy is avoided by applying our measurements to a specific single-domain state. The combination with a symmetry-based phenomenological theory that we develop for quantifying IFE and ICME allows us to distinguish between the two effects and clarify the driving force exciting the magnon oscillations in NiO. Moreover, we compare the magnon generation efficiencies of the two effects.

The paper is organized  as follows: the crystallographic and magnetic lattices of NiO are reviewed in Section~\ref{sec:NiOStructure} with a special focus on the domain structure. We describe the magneto-optical properties in Section~\ref{sec:theory1}. Subsequently, based on that description, we develop a theory for the inverse magneto-optical effects in NiO in Section~\ref{sec:theory2}. In Section~\ref{sec:Exp1}, the optical pump-probe setup is described, and the results of the theory sections are converted into experimental configurations that enable IFE and ICME to be measured and distinguished. Sections~\ref{sec:lin} and \ref{sec:circ} present the experimental results obtained by linear and circular pump polarizations, respectively. They are discussed in detail in Section~\ref{sec:Discussion}, where we show that magnon excitation via the ICME in NiO is significantly more efficient than via the IFE. In Section \ref{sec:conclusions}, conclusions are presented.

\subsection{NiO structure}\label{sec:NiOStructure}

\begin{figure}[b]
	\includegraphics[width = \columnwidth]{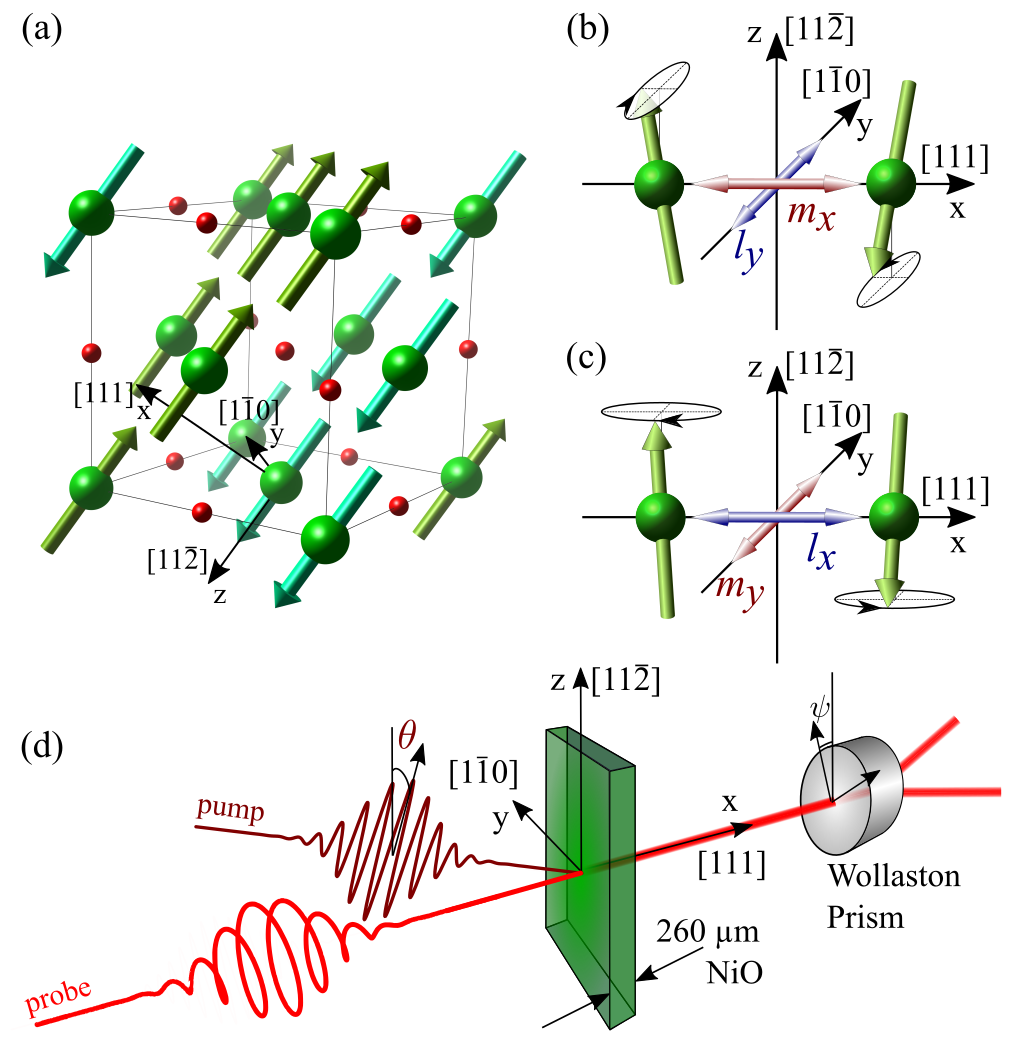}
	\caption{\label{fig:angles} (a) Crystallographic and magnetic structure of NiO in the defined coordinate system. (b) Graphical representation of the spin dynamics for the in-plane mode and (c) out-of-plane mode. (d) Schematic of the experimental geometry. $\theta$ denotes the azimuth angle of the pump polarization relative to the easy-axis of the spins, whereas $\psi$ parameterizes the setting of the Wollaston prism. The probe pulse is always circularly polarized.}
\end{figure}

NiO is a type-II antiferromagnet with a N\'eel temperature $T_N$ of \SI{523}{K}.\cite{Kondoh60} In the paramagnetic phase, the crystal has the NaCl-type structure (point group $m\bar{3}m$). Below $T_N$, spins are coupled ferromagnetically within the $\left\lbrace 111 \right\rbrace$ planes with neighboring planes being coupled antiferromagnetically [Fig.~\ref{fig:angles}(a)].\cite{RothSlack60} Furthermore, in the antiferromagnetic phase, there is a rhombohedral distortion along the $\left\langle 111 \right\rangle$ direction arising from exchange striction. This distortion corresponds to a reduction of the crystallographic point symmetry to $\bar{3}m$ and induces a significant uniaxial optical anisotropy of $\Delta n = 0.003$.\cite{Roth60} The optic axis forms along the direction of the distortion. Because the four independent $\left\langle111\right\rangle$ directions ($[111]$, $[11\bar{1}]$, $[1\bar{1}1]$, $[\bar{1}11]$) are energetically degenerate in the paramagnetic phase, the rhombohedral distortion can occur along any of those directions leading to four twin-domain states commonly referred to as $T$-domain states ($T_0$--$T_3$). The four $T$-domain states can be distinguished by their linear birefringence.\cite{Satoh10}

Within each $T$-domain state, spins point in one of three independent $\left\langle11\bar{2}\right\rangle$ directions that are perpendicular to the direction of the rhombohedral distortion.\cite{Hutchings72} This creates the formation of three spin domain states, commonly referred to as $S$-domain states, $S_1$--$S_3$, leading to a total of twelve possible orientation domain states in NiO.\cite{Sanger06} The formation of the $S$-domains leads to another small magnetostrictive distortion, corresponding to a reduced crystallographic point symmetry $2/m$, which is also the point symmetry of the magnetic lattice.\cite{Cracknell69,Fiebig01} This distortion, as well as the resulting linear birefringence, are approximately two orders of magnitude smaller than that associated with the $T$-domains\cite{KondohTakeda64} so that they have negligible influence on the polarization of the propagating pump and probe light. 
For the symmetry-based polarization analysis, however, the full magnetic $2/m$ symmetry needs to be considered, as we shall see later.
Antiferromagnetic ordering along the $\left[11\bar{2}\right]$ direction breaks the threefold rotational symmetry; for the resulting $2/m$ symmetry the twofold axis is perpendicular to both the rhombohedral distortion and the easy-axis of the spins, i.e., along $\left[1\bar{1}0\right]$.

With the two sublattice magnetizations $\mathbf{M}_1(t)$ and $\mathbf{M}_2(t)$, we define the ferromagnetic vector $\mathbf{M}(t) = \mathbf{M}_1(t)+\mathbf{M}_2(t)$ and the antiferromagnetic vector $\mathbf{L}(t) = \mathbf{M}_1(t)-\mathbf{M}_2(t)$. To study dynamics, it is convenient to split both quantities into a time-independent ground state and describe the excitation by a time-dependent contribution.
\begin{subequations}
\begin{equation}
	\mathbf{M}(t) = \mathbf{M}_0 + \mathbf{m}(t) = \mathbf{m}(t)
	\label{subeqn:Mt}
\end{equation}
\begin{equation}
	\mathbf{L}(t) = \mathbf{L}_0 + \mathbf{l}(t)
	\label{subeqn:Lt}
\end{equation}
\end{subequations}

The dynamic contribution may be a superposition of the two eigenmodes of the two sublattice antiferromagnetic system, both of which are optically excitable in NiO.\cite{SatohPRL10} For the in-plane mode (IPM) or $B_g$ mode, the modulation of the antiferromagnetic vector $\mathbf{l}(t)$ is along the $\left[1\bar{1}0\right]$ direction, i.e., it occurs within the sheets of ferromagnetically coupled spins. The oscillating magnetization $\textbf{m}(t)$, in contrast, is along the $\left[111\right]$ out-of-plane direction. The frequency of this mode is $\Omega_{\text{IPM}}/2\pi\simeq\SI{0.14}{THz}$ at \SI{77}{K}.\cite{Milano04,SatohPRL10} The opposite behavior occurs for the out-of-plane mode (OPM) or $A_g$ mode. The antiferromagnetic vector is modulated along the $\left[111\right]$ direction, whereas the magnetization oscillates along $\left[1\bar{1}0\right]$. The eigenfrequency of the out-of-plane mode is $\Omega_{\text{OPM}}/2\pi\simeq\SI{1.07}{THz}$ at \SI{77}{K}.\cite{Kondoh60, Grimsditch98, SatohPRL10, Nishitani10, Kampfrath11, Kanda11, Higuchi11, Nishitani12, Takahara12, Baierl16}

In contrast to previous publications,\cite{SatohPRL10} we specifically consider a $T_0$ domain on a (111)-cut NiO sample, where the rhombohedral distortion is along the surface normal. Therefore, the optic axis coincides with the propagation direction of light at normal incidence and optical anisotropy, especially linear birefringence, can be avoided. For this situation, we define a reference system: We choose the $x$-axis to be along the surface normal, i.e., the $\left[111\right]$-direction, the $z$-axis to be along the magnetic easy-axis, i.e., the $\left[11\bar{2}\right]$-direction, and the $y$-axis perpendicular to both to form a right-handed coordinate system, i.e., along $\left[1\bar{1}0\right]$. The orientation is shown in Fig.~\ref{fig:angles} together with a schematic representation of the spin motion for the in-plane mode [Fig.~\ref{fig:angles}(b)] and the out-of-plane mode [Fig.~\ref{fig:angles}(c)]. Using this notation, Eqs.~(\ref{subeqn:Mt}) and (\ref{subeqn:Lt}) can be expressed explicitly as
\begin{subequations}
	\begin{equation}
	\mathbf{M}(t) = \begin{pmatrix} m_x(t)\\ m_y(t) \\0 \end{pmatrix}
	\label{subeqn:Mt2}
	\end{equation}
	\begin{equation}
	\mathbf{L}(t) = \begin{pmatrix} 0\\ 0\\L_z \end{pmatrix} + \begin{pmatrix} l_x(t)\\ l_y(t)\\0\end{pmatrix}.
	\label{subeqn:Lt2}
	\end{equation}
\end{subequations}

\noindent Here, $m_x$ and $l_y$ are contributions purely from the in-plane mode, whereas $m_y$ and $l_x$ originate from the out-of-plane mode.

\section{Phenomenological theory of magneto-optical and inverse magneto-optical effects in NiO}\label{sec:theory0}

We briefly review the phenomenological theory of the Faraday effect as well as the Cotton-Mouton effect, both of which are used to detect magnon oscillations in NiO. Furthermore, a phenomenological theory of the inverse magneto-optical effects, i.e., the IFE and the ICME, is presented, which enables the different magnon excitation mechanisms to be distinguished. These discussions are accompanied by a special consideration of the point-group symmetry of NiO.

Light-matter interaction is typically described by an interaction Hamiltonian which, in cgs  units, reads \cite{LandauLifshitz84}
\begin{equation}
  \mathcal{H}_{\rm int} = -\frac{\epsilon_{ij}(\mathbf{M},\mathbf{L})}{16\pi}
\mathcal{E}_i(t)\mathcal{E}^\ast_j(t).
\label{eqn:Hamiltonian}
\end{equation}

\noindent Here, \eij$(\textbf{M},\textbf{L})$ is the dielectric tensor, which is in general a complex function of $\textbf{M}$ and $\textbf{L}$, and $\mathcal{E}_{i}$ is the electric field amplitude with $E_i(t) = \Re\left[\mathcal{E}_{i}(t)e^{i\omega t}\right]$.\cite{Kalashnikova08,Satoh15} We assume light propagating in $x$-direction. Thus, $\mathcal{E} = \left(0,\mathcal{E}_y(t),\mathcal{E}_z(t)\right)$.

Expanding the dielectric tensor into a power series in $\textbf{M}$ and $\textbf{L}$, we obtain with magneto-optical coupling constants $k_{ijk}$ and $g_{ijkl}$:\cite{Cottam86,Eremenko92}
\begin{eqnarray}
  \epsilon_{ij}
  &=& \epsilon^{(0)}_{ij}+ik^{\rm M}_{ijk}M_k+ik^{\rm L}_{ijk}L_k\nonumber\\
  &+&g^{\rm MM}_{ijkl}M_kM_l+g^{\rm LL}_{ijkl}L_kL_l+g^{\rm ML}_{ijkl}M_kL_l.
  \label{eqn:epsexpansion1}
\end{eqnarray}

\noindent As $\left|\textbf{M}\right|\ll\left|\textbf{L}\right|$, the term quadratic in $\textbf{M}$ can be neglected. For symmetry reasons, only even orders in $\textbf{L}$ can give non-vanishing contributions to the dielectric tensor. This leads to the simplified equation
\begin{equation}
  \epsilon_{ij}=\epsilon^{(0)}_{ij}+ik^{\rm M}_{ijk}M_k+g^{\rm LL}_{ijkl}L_kL_l.
\label{eqn:epsexpansion2}
\end{equation}

In the following discussions, the superscripts $\rm M$ and $\rm LL$ will be omitted.
\noindent Considering the complex dielectric tensor \eij$(\textbf{M},\textbf{L})$ and the Onsager principle, the absence of absorption leads to
\begin{equation}
  \epsilon_{ij}(\textbf{M},\textbf{L}) = \epsilon^\ast_{ji}(\textbf{M},\textbf{L})
= \epsilon^\ast_{ij}(-\textbf{M},\textbf{L})= \epsilon^\ast_{ij}(\textbf{M},-\textbf{L}).
\label{eqn:Onsager}
\end{equation}

\noindent Here, $\epsilon^\ast_{ij}$ denotes the complex conjugate of $\epsilon_{ij}$. Eq.~(\ref{eqn:Onsager}) indicates that the diagonal components $\epsilon_{ii}$ are purely real, whereas the off-diagonal components are in general complex. $\Re{\left[\epsilon_{ij}\right]}$ is a symmetric tensor, whereas $\Im{\left[\epsilon_{ij}\right]}$ is antisymmetric. Consequently, the nonzero coefficients in Eq.~(\ref{eqn:epsexpansion2}) are real-valued and satisfy $k_{ijk}=-k_{jik}$ and $g_{ijkl}=g_{jikl}=g_{ijlk}=g_{jilk}$. As the birefringence caused by the magnetostriction is neglected in our symmetry analysis,\cite{KondohTakeda64} we set $\epsilon^{(0)}_{yy}=\epsilon^{(0)}_{zz} \equiv \epsilon^{(0)}$ and $\epsilon^{(0)}_{yz}=\epsilon^{(0)}_{zy} \equiv 0$. Considering an electromagnetic wave propagating in $x$-direction, we neglect all $x$-components of the dielectric tensor and assume the following ansatz for the remaining tensor components
\begin{equation}
\begin{pmatrix}
\epsilon_{yy} & \epsilon_{yz} \\
\epsilon_{zy} & \epsilon_{zz}
\end{pmatrix}
=
\begin{pmatrix}
\epsilon^{(0)}+\widetilde{\alpha}_y & \beta+i\xi \\
\beta-i\xi & \epsilon^{(0)}+\widetilde{\alpha}_z
\end{pmatrix}.
\label{eqn:eps}
\end{equation}

\noindent With Eq.~(\ref{eqn:epsexpansion2}) we identify
\begin{subequations}
	\begin{eqnarray}
	\widetilde{\alpha}_y &=& g_{yyzz}L_zL_z+g_{yyzx}L_zl_x\label{eqn:7a}\\
	\widetilde{\alpha}_z &=& g_{zzzz}L_zL_z+g_{zzzx}L_zl_x\label{eqn:7b}\\
	\beta &=& g_{yzzy}L_zl_y\\
	\xi &=& k_{yzx}m_x.
	\end{eqnarray}
\end{subequations}

\noindent All other possible contributions to $g_{ijkl}$ and $k_{ijk}$ vanish in compliance with the $2/m$ symmetry of the antiferromagnetic order.\cite{Cracknell69,Eremenko92} As the static magnetic linear birefringence expressed by Eqs.~(\ref{eqn:7a}) and (\ref{eqn:7b}) was not resolved, we assume $g_{yyzz} \approx g_{zzzz}$ and redefine:
\begin{subequations}
	\begin{eqnarray}
	\epsilon^{(0)}+\widetilde{\alpha}_y &=& \epsilon^\prime+\alpha_y\\
	\epsilon^{(0)}+\widetilde{\alpha}_z &=& \epsilon^\prime+\alpha_z,
	\end{eqnarray}
\end{subequations}

\noindent with
\begin{subequations}
	\begin{eqnarray}
	\alpha_y &=& g_{yyzx}L_zl_x\\
	\alpha_z &=& g_{zzzx}L_zl_x.
	\end{eqnarray}
\end{subequations}

\subsection{Magneto-optical effects}\label{sec:theory1}

We now discuss the eigenvalues and eigenpolarizations of Eq.~(\ref{eqn:eps}) in the simplified case, where only one of the quantities $\alpha$, $\beta$ or $\xi$ is non-zero. The square roots of these eigenvalues are the refractive indices corresponding to the eigenpolarizations. We show that $\xi$ leads to the Faraday effect, i.e., circular birefringence, whereas $\alpha$ and $\beta$ induce a linear birefringence thus leading to the Cotton-Mouton effect.

\subsubsection{$\xi\neq0, \alpha=\beta=0$}

The refractive indices $N_\pm$ and corresponding eigenpolarizations $\textbf{E}_\pm$ are
\begin{subequations}
	\begin{equation}
	N_\pm = \sqrt{\epsilon^\prime\pm\xi}
	\label{subeqn:gamma2}
	\end{equation}
	\begin{equation}
	\textbf{E}_\pm = \frac{E_0}{\sqrt{2}}\exp\left\lbrace i\omega\left(t-\frac{N_\pm}{c} x\right) \right\rbrace\left(\mathbf{\hat{y}}\mp i\mathbf{\hat{z}}\right).
	\label{subeqn:gamma1}
	\end{equation}
\end{subequations}

\noindent Here, $\mathbf{\hat{y}}$ and $\mathbf{\hat{z}}$ correspond to unit vectors along the $y$- and $z$-directions, $\omega$ is the angular frequency of the light, and $c$ is the speed of light. Thus, the eigenpolarizations $\textbf{E}_\pm$ describe circularly polarized waves ($\sigma^\pm$), which are subject to different refractive indices $N_\pm$. Typically, with $\epsilon^\prime \gg \lvert k_{yzx}m_x \rvert$, the circular birefringence $\Delta N = N_+ - N_- \approx k_{yzx}m_x/\sqrt{\epsilon^\prime}$ is linear in $m_x$ and results in a rotation of the plane of polarization of linearly polarized light by
\begin{equation}
  \phi_F = -\frac{\omega d k_{yzx}m_x}{c\sqrt{\epsilon^\prime}} \propto m_x,
\label{eqn:mxphase}
\end{equation}

\noindent where $d$ is the sample thickness. Therefore, the magnetization component $m_x$ can be studied by analyzing the Faraday rotation of linearly polarized light.

\subsubsection{$\beta\neq0, \alpha=\xi=0$} The eigenvalues $N_{\pm45^{\circ}}$ and the corresponding eigenpolarizations $\textbf{E}_{\pm45^{\circ}}$ are:
\begin{subequations}
	\begin{equation}
	N_{\pm45^{\circ}} = \sqrt{\epsilon^\prime\pm \beta}
	\label{subeqn:beta1}
	\end{equation}
	\begin{equation}
	\textbf{E}_{\pm45^{\circ}} = \frac{E_0}{\sqrt{2}}\exp\left\lbrace i\omega\left(t-\frac{N_{\pm45^{\circ}}}{c} x\right) \right\rbrace\left(\mathbf{\hat{y}}\pm\mathbf{\hat{z}}\right).
	\label{subeqn:beta2a}
	\end{equation}
\end{subequations}

\noindent The eigenpolarizations are linearly polarized with angle $\pm\,45^\circ$ relative to the $y$-direction. Over a propagation distance $d$, this linear birefringence induces a phase difference of
\begin{equation}
  \phi_{45^\circ} \approx
\frac{\omega d g_{yzzy}L_zl_y}{c\sqrt{\epsilon^\prime}}\propto l_y.
\label{eqn:lyphase}
\end{equation}

\noindent Incident circularly polarized light thus becomes elliptically polarized with principal axes along the $y$- and $z$-directions.

\subsubsection{$\alpha\neq0, \beta=\xi=0$}

The refractive indices $N_{y,z}$ and eigenpolarizations $\textbf{E}_{y,z}$ are:
\begin{subequations}
	\begin{equation}
	N_{y,z} = \sqrt{\epsilon^\prime+\alpha_{y,z}}
	\label{subeqn:delta1}
	\end{equation}
	\begin{eqnarray}
	\textbf{E}_{y} &= E_0\exp\left\lbrace i\omega\left(t-\frac{N_{y}}{c} x\right)\right\rbrace\mathbf{\hat{y}}\nonumber\\
	\textbf{E}_{z} &= E_0\exp\left\lbrace i\omega\left(t-\frac{N_{z}}{c} x\right)\right\rbrace\mathbf{\hat{z}}.
	\label{subeqn:delta2a}
	\end{eqnarray}
\end{subequations}

\noindent Hence, the eigenpolarizations are linearly polarized along the $y$- and $z$-direction with different refractive indices $N_{y,z}$. Over a propagation distance $d$, this linear birefringence induces a phase difference of
\begin{equation}
  \phi_{yz} \approx \frac{\omega d (g_{yyzx}-g_{zzzx})L_zl_x}{2c\sqrt{\epsilon^\prime}}\propto l_x.
\label{eqn:lxphase}
\end{equation}

\noindent Consequently, circularly polarized light becomes elliptically polarized with principal axes aligned at $\pm\,45^{\circ}$.
The magnetically induced linear birefringence observed in cases 2 and 3 are also known collectively as the Cotton-Mouton effect.

To summarize, because each component of the dielectric tensor has a specific dynamical modification, the polarization of light propagating through the material is altered in a highly selective way. This selectivity enables the different physical mechanisms that are responsible for a certain modulation of the magnetization to be distinguished experimentally. In particular, only the in-plane magnon mode causes oscillations in $m_x$ and $l_y$ and can thus be observed via the Faraday effect (case 1) or the Cotton-Mouton effect (case 2). The out-of-plane magnon mode causes oscillations of $l_x$ and $m_y$ and is therefore only observable via the Cotton-Mouton effect (case 3).

\subsection{Inverse magneto-optical effects}\label{sec:theory2}

Based on the interaction Hamiltonian defined in Eq.~(\ref{eqn:Hamiltonian}) and the exact same dielectric tensor defined in Eq.~(\ref{eqn:eps}) discussed in regard to the crystal symmetry specific to NiO, we can also describe the inverse magneto-optical effects. In accordance with the previous notion, a magneto-optical coupling via $k_{ijk}$ leads to the IFE, whereas coupling via $g_{ijkl}$ represents the ICME.

We define the effective magnetic fields $\textbf{H}^\mathrm{eff}$ and $\textbf{h}^\mathrm{eff}$ for $\textbf{m}$ and $\textbf{l}$ as the partial derivative of the interaction Hamiltonian with respect to $\textbf{m}$ and $\textbf{l}$:
\begin{eqnarray}
\textbf{H}^\mathrm{eff} = -\frac{\partial\mathcal{H}_{\rm int}}{\partial\textbf{m}} &\hspace{1.5cm}& \textbf{h}^\mathrm{eff} = -\frac{\partial\mathcal{H}_{\rm int}}{\partial\textbf{l}}.
\label{eqn:fields}
\end{eqnarray}
When an ultrashort light pulse irradiates a sample, these effective magnetic fields become the driving force of the non-thermal magnetization dynamics.

The Landau-Lifshitz-Gilbert equations for $\mathbf{m}$ and $\mathbf{l}$ are\cite{LandauLifshitz84}
\begin{subequations}
	\begin{eqnarray}
	\frac{d\mathbf{m}}{dt} &=&-\gamma\left\lbrace\mathbf{M}\times\mathbf{H}^\mathrm{eff}+\mathbf{L}\times\mathbf{h}^\mathrm{eff}\right\rbrace+\mathbf{R}_m
	\end{eqnarray}
	\begin{eqnarray}
	\frac{d\mathbf{l}}{dt} &=&-\gamma\left\lbrace\mathbf{M}\times\mathbf{h}^\mathrm{eff}+\mathbf{L}\times\mathbf{H}^\mathrm{eff}\right\rbrace+\mathbf{R}_l,
	\end{eqnarray}
\end{subequations}

\noindent where $\gamma$ is the gyromagnetic ratio. Anisotropy terms leading to the elliptical precession of $\mathbf{m}$ and $\mathbf{l}$, and damping terms are subsumed into $\mathbf{R}_{m,l}$. Combining these with Eq.~(\ref{eqn:fields}) and the initial conditions $\mathbf{M}(t=0) = 0$ and $\mathbf{L}(t=0) = (0,0,L_z)$, we obtain
\begin{subequations}
	\begin{eqnarray}
	\frac{d\mathbf{m}}{dt}
&=&\frac{\gamma}{16\pi}L^2_z[g_{yzzy}
\left\lbrace\mathcal{E}_y(t)\mathcal{E}^\ast_z(t)
+\mathcal{E}_z(t)\mathcal{E}^\ast_y(t)\right\rbrace\mathbf{\hat{x}}\nonumber\\
&-& \left\lbrace g_{yyxz}\mathcal{E}_y(t)\mathcal{E}^\ast_y(t)
+g_{zzxz}\mathcal{E}_z(t)\mathcal{E}^\ast_z(t)\right\rbrace\mathbf{\hat{y}}]
+\mathbf{R}_m\nonumber\\
	\label{eqn:LLGm}
	\end{eqnarray}
	\begin{eqnarray}
	\frac{d\mathbf{l}}{dt}
&=&-\frac{i\gamma}{16\pi}L_zk_{yzx}\left\lbrace\mathcal{E}_y(t)\mathcal{E}^\ast_z(t)-\mathcal{E}_z(t)\mathcal{E}^\ast_y(t)\right\rbrace\mathbf{\hat{y}}+\mathbf{R}_l.\nonumber\\
	\label{eqn:LLGl}
	\end{eqnarray}
\end{subequations}

If the magnetization dynamics are induced by an ultrafast laser pulse, which is short compared with the spin oscillation period, i.e., $\mathcal{E}(t)\mathcal{E}^\ast(t) \approx I_0\delta(t)$, the terms $\mathbf{R}_m$ and $\mathbf{R}_l$ can be neglected and Eqs.~(\ref{eqn:LLGl}) and (\ref{eqn:LLGm}) can be integrated around $t=0$:
\begin{subequations}
	\begin{eqnarray}
	\Delta \mathbf{m} &= &\frac{\gamma}{16\pi}L^2_z
[g_{yzzy}\left\lbrace\mathcal{E}_y\mathcal{E}^\ast_z
+\mathcal{E}_z\mathcal{E}^\ast_y\right\rbrace\mathbf{\hat{x}}\nonumber\\
&-& \left\lbrace g_{yyxz}\mathcal{E}_y\mathcal{E}^\ast_y
+g_{zzxz}\mathcal{E}_z\mathcal{E}^\ast_z\right\rbrace\mathbf{\hat{y}}]
	\label{eqn:LLGmint}
	\end{eqnarray}
	\begin{eqnarray}
	\Delta \mathbf{l} &=& -i\frac{\gamma}{16\pi}L_zk_{yzx}
\left\lbrace\mathcal{E}_y\mathcal{E}^\ast_z-\mathcal{E}^\ast_y\mathcal{E}_z\right\rbrace\mathbf{\hat{y}}.
	\label{eqn:LLGlint}
	\end{eqnarray}
\end{subequations}

\noindent These optically induced changes occur instantaneously during the excitation.

\subsubsection{Excitation by linearly polarized light}

With linearly polarized light specified by $(\mathcal{E}_y(t),\mathcal{E}_z(t))\,=\,\mathcal{E}(t)(\sin\theta,\cos\theta)$, where $\theta$ denotes the angle between the direction of polarization and the $z$-axis (cf. \ Fig.~\ref{fig:angles}), Eqs.~(\ref{eqn:LLGmint}) and (\ref{eqn:LLGlint}) lead to
\begin{subequations}
	\begin{eqnarray}
	\Delta \mathbf{m}^\mathrm{lin} =\frac{\gamma}{16\pi}
&L^2_zI_0[&g_{yzzy}\sin(2\theta)\mathbf{\hat{x}}\nonumber\\
&-&\left(g_1-g_2\cos(2\theta)\right)\mathbf{\hat{y}}]
	\end{eqnarray}
	\begin{eqnarray}
	\Delta \mathbf{l}^\mathrm{lin} = 0.
	\end{eqnarray}
\end{subequations}

\noindent Here, $g_1 = (g_{yyxz}+g_{zzxz})/2$ and $g_2 = (g_{yyxz}-g_{zzxz})/2$. After the quasi-instantaneous generation of $m_x$ and $m_y$, the spins start to precess around their easy-axis orientation with a strong ellipticity that reflects the pronounced magnetic anisotropy perpendicular to this axis. The short axis of the ellipse is along $\Delta\mathbf{m}$, whereas the long axis is along $\Delta\mathbf{l}$.\cite{Rubano10,Satoh10} The precession can be separated into in-plane and out-of-plane contributions, where for the in-plane mode $m_x$ and $l_y$ oscillate with a $\pi/2$ phase difference at frequency $\Omega_\mathrm{IPM}$ and for the out-of-plane mode $m_y$ and $l_x$ oscillate at frequency $\Omega_\mathrm{OPM}$. The magnetization dynamics lead to
\begin{subequations}
	\begin{eqnarray}
	m^\mathrm{lin}_x(t) &=& \hspace{0.3cm} \frac{\gamma}{16\pi}L^2_zI_0
g_{yzzy}\sin2\theta\cos\Omega_\mathrm{IPM}t\label{eqn:mlinIPM}\\
	l^\mathrm{lin}_y(t) &=& -\frac{\gamma}{16\pi}A_\mathrm{IPM}L^2_zI_0
g_{yzzy}\sin2\theta\sin\Omega_\mathrm{IPM}t
	\label{eqn:llinIPM}\\
	m^\mathrm{lin}_y(t) &=& -\frac{\gamma}{16\pi}L^2_zI_0
\left(g_1-g_2\cos2\theta\right)\cos\Omega_\mathrm{OPM}t\nonumber\\
	\\
	l^\mathrm{lin}_x(t) &=& -\frac{\gamma}{16\pi}A_\mathrm{OPM}L^2_zI_0
\left(g_1-g_2\cos2\theta\right)\sin\Omega_\mathrm{OPM}t.\nonumber\\
	\label{eqn:llinOPM}
	\end{eqnarray}
\end{subequations}

\noindent Here, we introduced the anisotropy factors $A_\mathrm{IPM}$ and $A_\mathrm{OPM}$, which account for the magnetic anisotropy and parameterize the ellipticity of the spin precession.\cite{Kalashnikova08} The coupling between the light field and the magnetization is purely described by parameters based on the tensor $g_{ijkl}$, and therefore based on magnetic linear birefringence. Therefore, both modes are excited by the ICME.

\subsubsection{Excitation by circularly polarized light}

With circularly polarized light, $\sigma^\pm = (\mathcal{E}_y(t),\mathcal{E}_z(t))\,=\,\mathcal{E}(t)(1,\mp i)/\sqrt{2}$, analogous considerations as for linearly polarized light lead to
\begin{subequations}
	\begin{eqnarray}
	\Delta \mathbf{m}^{\sigma^\pm}
&=& -\frac{\gamma}{16\pi}L^2_zI_0g_1\mathbf{\hat{y}}
	\label{eqn:LLGmintlin}
	\end{eqnarray}
	\begin{eqnarray}
	\Delta \mathbf{l}^{\sigma^\pm}
&=& \mp\frac{\gamma}{16\pi}L_zI_0k_{yzx}\mathbf{\hat{y}}.
	\label{eqn:LLGlintlin}
	\end{eqnarray}
\end{subequations}

\noindent This induces oscillations of $\mathbf{m}$ and $\mathbf{l}$ according to

\begin{subequations}
	\begin{eqnarray}
	m^{\sigma^\pm}_x(t) &=& \mp\frac{\gamma}{16\pi} \frac{1}{A_\mathrm{IPM}} L_zI_0 k_{yzx} \sin\Omega_\mathrm{IPM}t \label{eqn:mcircIPM}\\
	l^{\sigma^\pm}_y(t) &=& \mp\frac{\gamma}{16\pi}L_zI_0 k_{yzx} \cos\Omega_\mathrm{IPM}t,\label{eqn:lcircIPM}\\
	m^{\sigma^\pm}_y(t) &=& -\frac{\gamma}{16\pi} L^2_zI_0g_1 \cos\Omega_\mathrm{OPM}t\\
	l^{\sigma^\pm}_x(t) &=& -\frac{\gamma}{16\pi}A_\mathrm{OPM}L^2_zI_0g_1 \sin\Omega_\mathrm{OPM}t.\label{eqn:lcircOPM}
	\end{eqnarray}
\end{subequations}

\noindent Thus, the in-plane mode is linearly dependent on $L_z$, obtains a $180^\circ$ phase change upon changing the pump helicity, and couples via $k_{yzx}$, which is related to magnetic circular birefringence. Accordingly, it is excited by the IFE, which creates an effective magnetic field that exerts a torque on the spin system and contributes the term $\Delta\mathbf{l}$. Meanwhile, even though induced by circularly polarized light, the out-of-plane mode is excited via the ICME.

\section{Experimental results}

\subsection{Optical setup}\label{sec:Exp1}

We study the magnon dynamics in NiO by performing impulsive stimulated Raman scattering experiments in the time domain, which was realized by a pump-probe setup. We optically excite the sample using a 0.98-eV, 90-fs laser pulse and probe the transient optical properties of the material with a 1.55-eV, 50-fs pulse.\cite{SatohPRL10} The absorption coefficient of NiO for the pump pulse is approximately \SI{20}{\per\centi\metre} at \SI{77}{K}.\cite{Newman59} By pumping and probing the sample in the highly transparent regime, we are able to excite and measure the entire volume of our 260-$\mu$m thick NiO slice. Furthermore, we avoid heating effects, which allows us to study the non-thermal magnetization dynamics. The polarization of both pulses can be tuned such that any linear or circular polarization can be realized for the pump and for the probe pulse. The transmitted part of the probe pulse is split into orthogonal contributions by a Wollaston prism and measured as intensities $I_1$ and $I_2$ on a balanced pair of photodiodes. The theory presented in the previous section allows to predict the resulting imbalance 
\begin{equation}
\Delta\eta=\left[\frac{I_1-I_2}{I_1+I_2}\right]_{\rm{pump~on}}-\left[\frac{I_1-I_2}{I_1+I_2}\right]_{\rm{pump~off}}
\end{equation}
 between the photodiodes as a function of the orientation of the Wollaston prism, which is parameterized by the angle $\psi$ [cf. Fig.~\ref{fig:angles}(c)], as well as by the pump and probe polarizations. We focus on the Cotton-Mouton effect by probing with circularly polarized light and measuring the ellipticity of the transmitted light. This enables both in-plane and out-of-plane modes to be observed. The sample is kept at \SI{77}{K} for all measurements.

Eliminating $l_y$ by combining Eqs.~(\ref{eqn:lyphase}) and (\ref{eqn:llinIPM}), we find for the in-plane mode excited by linearly polarized light the following dependence of the ellipticity on the pump and probe conditions:
\begin{equation}
  \Delta\eta^\mathrm{lin}_\mathrm{IPM} =\mathcal{C}A_\mathrm{IPM}L^3_zI_0
g^{\rm pu}_{yzzy}g^{\rm pr}_{yzzy}\sin2\theta\cos2\psi\sin\Omega_\mathrm{IPM}t.
\label{eqn:prediction1}
\end{equation}

\noindent Here, we defined $\mathcal{C}=-\gamma\omega d/(16\pi c\sqrt{\epsilon^\prime})$. Furthermore, the magneto-optical coupling constants are in general frequency dependent and can therefore be different for the pump and the probe pulse. This is taken into account by introducing $g^{\rm pu}_{yzzy}$ and $g^{\rm pr}_{yzzy}$.

Analogously, combining Eqs.~(\ref{eqn:lyphase}) and (\ref{eqn:lcircIPM}) yields the following dependence for the observation of the in-plane mode, when excited by circularly polarized light:
\begin{equation}
  \Delta\eta^{\sigma^\pm}_\mathrm{IPM} = \pm\mathcal{C}L^2_zI_0 k^{\rm pu}_{yzx}g^{\rm pr}_{yzzy}\cos2\psi\cos\Omega_\mathrm{IPM}t.
\label{eqn:prediction2}
\end{equation}

Similar considerations based on Eqs.~(\ref{eqn:lxphase}) and (\ref{eqn:llinOPM}) as well as (\ref{eqn:lcircOPM}) yield for the out-of-plane mode:
\begin{eqnarray}
  \Delta\eta^\mathrm{lin}_\mathrm{OPM}
&=&\mathcal{C}A_\mathrm{OPM} L^3_z I_0\nonumber\\&\times& 
(g^{\rm pu}_1-g^{\rm pu}_2\cos2\theta)g^{\rm pr}_2\sin2\psi\sin\Omega_\mathrm{OPM}t\label{eqn:prediction3}\\
  \Delta\eta^{\sigma^\pm}_\mathrm{OPM} &=& \mathcal{C}A_\mathrm{OPM} L^3_zI_0
g^{\rm pu}_1g^{\rm pr}_2\sin2\psi\sin\Omega_\mathrm{OPM}t
\label{eqn:prediction4}
\end{eqnarray}

\noindent Thus, the present model clearly predicts the measurable signal of the magnon dynamics as a function of pump and probe polarizations. In reverse, it allows the determination of the mechanisms leading to magnon excitation. Experimentally verifying the predictions, which are ultimately summarized in Eqs.~(\ref{eqn:prediction1}) to (\ref{eqn:prediction4}), is the core part of the following section. We shall first consider excitations using linearly polarized pump pulses and subsequently circularly polarized light.

\subsection{Excitation by linearly polarized light}\label{sec:lin}

To verify the predictions regarding linearly polarized pump pulses, i.e., Eqs.~(\ref{eqn:prediction1}) and (\ref{eqn:prediction3}), we performed time-resolved measurements for three different settings:
\begin{enumerate}
	\item[i.] The detection angle $\psi$ is fixed at $0^\circ$ and the pump polarization angle $\theta$ is varied.
	\item[ii.] The pump polarization angle $\theta$ is fixed at $45^\circ$ and the detection angle $\psi$ is varied.
	\item[iii.] The detection angle $\psi$ is fixed at $45^\circ$ and the pump polarization angle $\theta$ is varied.
\end{enumerate}

\begin{figure}[b]
	\includegraphics[width = \columnwidth]{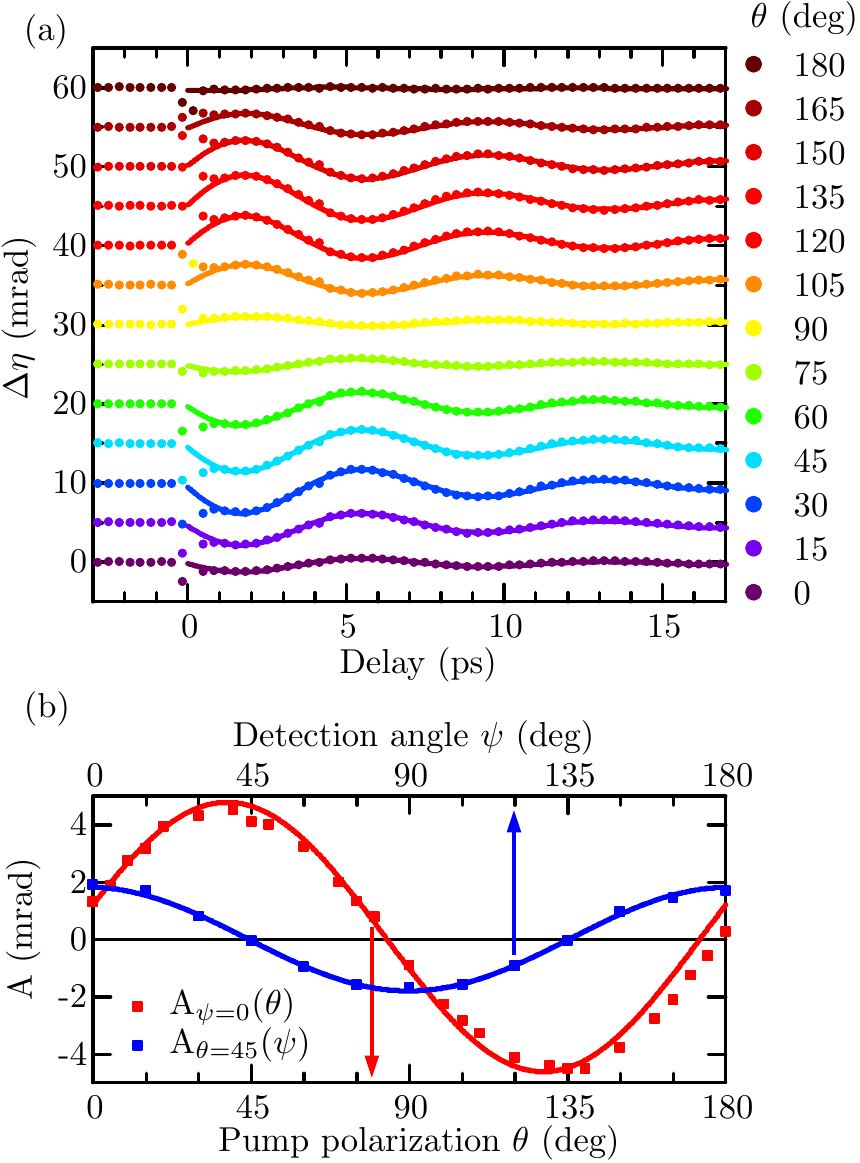}
	\caption{\label{fig:linIPM} (a) Observation of the in-plane mode from measurements of the Cotton-Mouton effect in setting (i), i.e., $\psi = 0^\circ$, $\theta$ varied. Curves are vertically displaced for clarity. (b) Signed amplitude $A$ of optically induced magnon oscillation in setting (i) (red) and setting (ii) (blue). $A(\theta)\propto \sin2(\theta+\zeta)$. $A(\psi)\propto \cos2\psi$. The difference in the modulation amplitude reflects the difference in pump power as mentioned in the text.}
\end{figure}

Figure \ref{fig:linIPM}(a) shows time-resolved ellipticity measurements for setting (i). The figure shows a single oscillation with a periodicity of approximately \SI{8}{ps}. The solid lines are fits according to the equation\ $\Delta\eta_\mathrm{IPM} = \eta_0-A\exp\left(-t/\tau\right)\sin\left(\Omega t + B\right)$. (Note that $A$ corresponds to $\mathcal{C}A_\mathrm{IPM}L^3_zI_0 g^{\rm pu}_{yzzy}g^{\rm pr}_{yzzy}\sin2\theta\cos2\psi$ in Eq.~(\ref{eqn:prediction1}).) Fitting yields an oscillation frequency $\Omega/2\pi$ of \SI{0.13}{THz}, which is in agreement with the expected value of \SI{0.14}{THz} for the in-plane mode.\cite{SatohPRL10} The slight deviation may be temperature related. The initial phase $B$ turns out to be close to zero, confirming the sine-like time-dependence of Eq.~(\ref{eqn:prediction1}). The red curve in Fig.~\ref{fig:linIPM}(b) shows the behavior of the signed amplitude $A$. It resembles the predicted $\sin2\theta$ function, but a fit proportional to $\sin2\left(\theta-\zeta\right)$ reveals a small shift $\zeta=-6.9^\circ\pm0.7^\circ$, and thus a deviation from the predicted behavior. As we shall see in Section \ref{sec:Sdomains}, this phase shift originates from the $S$-domain substructure of our single-$T$-domain. Distinct from the red curve, the blue curve in Fig.~\ref{fig:linIPM}(b) shows the signed amplitude $A$ of the magnon oscillation for setting (ii). It confirms the expected $\cos2\psi$ dependence of the in-plane mode amplitude in both Eqs.~(\ref{eqn:prediction1}) and (\ref{eqn:prediction2}). To verify the linear dependence on the pump intensity, the pump fluence was reduced from \SI{80}{mJ/cm^2} for setting (i) to \SI{40}{mJ/cm^2} for setting (ii). The observed maximum amplitudes of the two curves in Fig.~\ref{fig:linIPM}(b) differ by a factor of about 2, confirming the predicted behavior.

\begin{figure}[b]
	\includegraphics[width = \columnwidth]{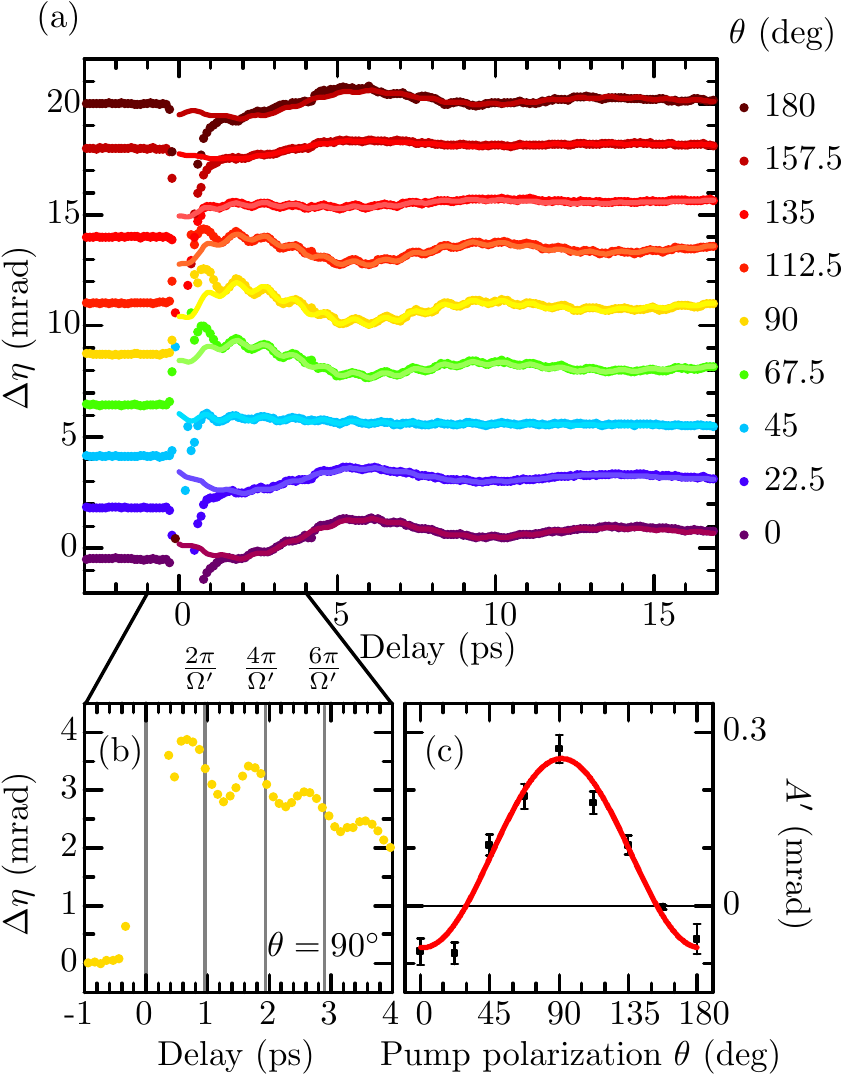}	
	\caption{\label{fig:linOPM}(a) Observation of the out-of-plane mode by measurement of the Cotton-Mouton effect in setting (iii), i.e., $\psi = 45^\circ$, $\theta$ varied. Curves are vertically displaced for clarity. (b) Magnification of region between \SI{-1}{ps} and \SI{4}{ps}. The out-of-plane mode is sine-like. Data points around 0 are out of scale. (c) Dependence of the signed amplitude $A^\prime$ of the out-of-plane mode on the linear pump polarization angle $\theta$. The solid red line is a fit using $A^\prime = S-T\cos2\left(\theta-\zeta\right)$.}
\end{figure}

Figure \ref{fig:linOPM}(a) shows time-resolved measurements of the magnetically induced linear birefringence for setting (iii). According to our model, this allows for the most efficient observation of the out-of-plane mode. Measurements were performed on the same spot as for Fig.~\ref{fig:linIPM}(a). A high-frequency modulation of the underlying in-plane mode is clearly visible. The solid curves are fits according to $\Delta\eta = \eta_0 + A_0\exp\left(-t/\tau_0\right) -A\exp\left(-t/\tau\right)\sin\left(\Omega t+B\right)-A^\prime \exp\left(-t/\tau^\prime\right) \sin\left(\Omega^\prime t+B^\prime\right)$. The exponential terms ($\sim\tau_0, \tau, \tau^{\prime}$) and the phase shifts ($\sim B, B^{\prime}$) are phenomenological additions parameterizing the magnetic damping and the aforementioned $S$-domain-substructure of our single-$T$-domain, respectively. The fit reveals $\Omega/2\pi = \SI{0.13}{THz}$ and $\Omega^\prime/2\pi = \SI{1.07}{THz}$ confirming the origin of the observed oscillations as a magnon excitation. A magnified representation of the region around $t = 0$ is given in Fig.~\ref{fig:linOPM}(b). The sine-like behavior of the out-of-plane mode is in agreement with Eq.~(\ref{eqn:prediction3}). Figure~\ref{fig:linOPM}(c) shows the dependence of the signed oscillation amplitude of the out-of-plane mode $A^\prime$ on the pump polarization angle $\theta$. As indicated by Eq.~(\ref{eqn:prediction3}), the red line plots the fitting function $X_1-X_2\cos2\left(\theta-\zeta\right)$ with $X_1 = (9\pm1)\times10^{-5}$, $X_2 = -(1.6\pm0.1)\times 10^{-4}$ and $\zeta = 3.7^\circ\pm 2.2^\circ$. The phase shift of $3.7^\circ$ and the presence of the in-plane mode are again caused by the admixture of additional $S$-domains to the anticipated single-domain state, which are discussed in detail in Section \ref{sec:Sdomains}.

Summarizing, we are able to observe both magnon modes of NiO by studying the magnetically induced linear birefringence, which can be efficiently probed by circularly polarized light. Furthermore, based on the striking agreement between measurement and theory, we can identify the ICME as the driving mechanism for the optical magnon excitation by linearly polarized light in NiO.

\subsection{Excitation by circularly polarized light}
\label{sec:circ}

\begin{figure}[b]
	\includegraphics[width = \columnwidth]{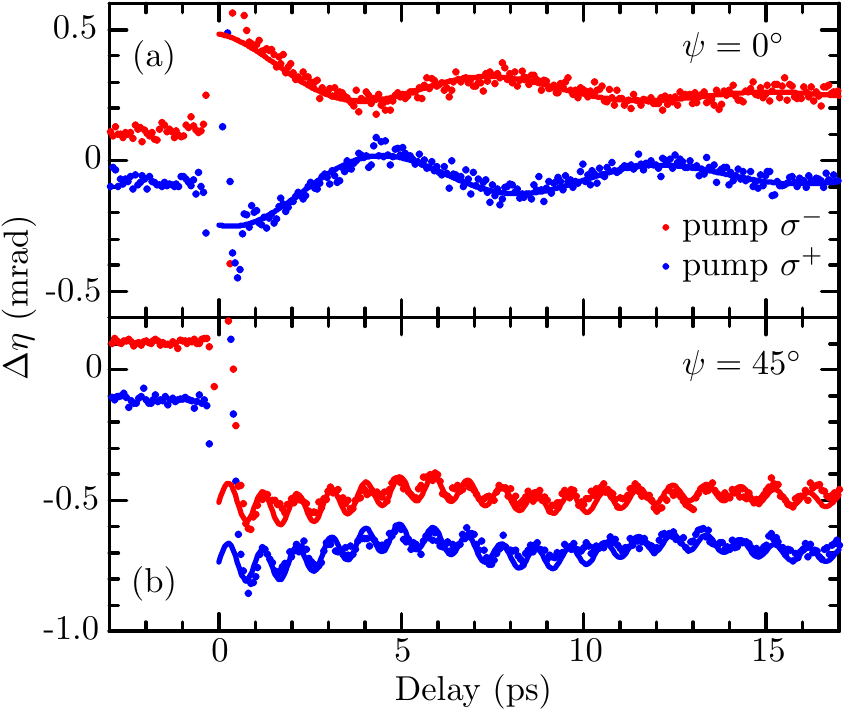}
	\caption{\label{fig:circ} Magnon oscillations induced by circularly polarized light. Curves have been displaced by $\pm\SI{0.1}{mrad}$ for clarity. (a) Only the in-plane mode is observed; the signal displays a $180^\circ$ phase shift following a change in the pump helicity, indicating excitation via the IFE. (b) The out-of-plane mode has no pump-helicity dependence, indicating excitation via the ICME.}
\end{figure}

After confirming our model theory for the generation of magnons by linearly polarized light, we now consider magnon excitations driven by circularly polarized optical pulses. Similar to the previous section, two cases can be distinguished, where the detection angle of the Wollaston prism is fixed to either $\psi = 0^\circ$ or $\psi = 45^\circ$. Furthermore, the helicity $\sigma^\pm$ of the circularly polarized pump pulse can be altered. Four individual measurements are obtained (see Fig.~\ref{fig:circ}).

For $\psi = 0^\circ$ [Fig.~\ref{fig:circ}(a)], only the in-plane mode is observed in agreement with the theory. The cosine-like behavior of the probed birefringence accords also with the prediction. Moreover, the in-plane mode obtains a $180^\circ$ phase shift when the pump helicity is changed. This is a distinct signature of the IFE as the driving mechanism of this oscillation. Microscopically, the IFE creates an effective magnetic field pulse in $x$-direction, which acts as a torque on $L_z$, effectively rotating $\mathbf{L}$ around the $x$-axis. This causes a finite contribution in $l_y$, which can be consequently probed by the induced birefringence via the Cotton-Mouton effect.

The out-of-plane mode cannot be probed in this geometry because of the $\sin2\psi$ dependence [Eqs.~(\ref{eqn:prediction3}) and (\ref{eqn:prediction4})]. To clarify its excitation mechanism, we also took measurements at $\psi = 45^\circ$, which allows for the observation of the out-of-plane mode. The obtained time-traces [Fig.~\ref{fig:circ}(b)] show the expected sine-like time-dependence. Remarkably, the out-of-plane mode does not obtain a $180^\circ$ phase shift after a change in the pump helicity, just as predicted by Eq.~(\ref{eqn:prediction4}). Consequently, based on the excellent agreement between theory and all measurements presented here, we can identify the Cotton-Mouton effect as the driving mechanism of the out-of-plane mode, even though it was excited by circularly polarized light. It is worth noting that the weak underlying signature of the in-plane mode in Fig.~\ref{fig:circ}(b) does not obtain a $180^\circ$ phase shift, when the pump helicity is changed. Furthermore, it exhibits a sine-like time-dependence as opposed to the cosine-like time-dependence of the in-plane mode in Fig.~\ref{fig:circ}(a). Thus, it is not excited by the IFE acting on the underlying $S$-domain substructure, but rather by the ICME [Eq.~(\ref{eqn:prediction1})] due to a slight inevitable ellipticity of the circularly polarized pump pulse.

\section{Discussion}
\label{sec:Discussion}

\subsection{Influence of $S_2$- and $S_3$-domains}\label{sec:Sdomains}

The coordinate system in Fig.~\ref{fig:angles} was chosen such that the easy-axis of the $S_1$-domain state lies along the $z$-axis. The probed single-$T$-domain area, however, may also include $S_2$- and $S_3$-domains. Their easy axes are rotated around the $x$-axis by $120^\circ$ and $240^\circ$, respectively. For a more detailed analysis of our data, we therefore have to expand $\Delta\eta$ by terms representing the contributions from these domain states. For pumping with \textit{linearly polarized light}, probing with circularly polarized light leads to
\begin{eqnarray}
\Delta\eta^\mathrm{lin}_\mathrm{IPM}
  &=&\mathcal{C}A_\mathrm{IPM}L^3_zI_0 g^{\rm pu}_{yzzy}g^{\rm pr}_{yzzy}\sin\Omega_\mathrm{IPM}t\nonumber\\ &&\times[A_1\cos\left(2\psi\right)\sin\left(2\theta\right)\nonumber\\
  &&+A_2\cos\left(2\psi-120^\circ\right)\sin\left(2\theta-120^\circ\right)\nonumber\\
  &&+A_3\cos\left(2\psi-240^\circ\right)\sin\left(2\theta-240^\circ\right)]
\label{eqn:S2S3}
\end{eqnarray}

\noindent for the in-plane mode, where $A_{1,2,3}$ represent the area fractions covered by the domain states $S_{1,2,3}$ of the single $T$-domain. Thus, we impose the boundary condition $A_1 + A_2 + A_3 = 1$. Note that for $A_1 = A_2 = A_3 = 1/3$, the isotropic $\bar{3}m$ symmetry is recovered as an average across the $T$-domain. In this case, Eq.~(\ref{eqn:S2S3}) simplifies to
\begin{equation}
  \Delta\eta = \frac{1}{2}\mathcal{C}A_\mathrm{IPM}L^3_zI_0g^{\rm pu}_{yzzy}g^{\rm pr}_{yzzy} \sin\left(2\theta-2\psi\right)\sin\Omega_\mathrm{IPM}t,
\label{eqn:Sisotropic}
\end{equation}

\noindent indicating that the observed amplitude depends solely on the difference between pump and detection angle. This behavior has been observed for instance in FeBO$_3$.\cite{Kalashnikova08}

For parameterizing the degree of $S$-domain mixing within a NiO $T$-domain, it is convenient to consider the setting $\psi = 0$ for which Eq.~(\ref{eqn:S2S3}) can be rewritten as
\begin{eqnarray}
  \Delta\eta^\mathrm{lin}_\mathrm{IPM} &=& \mathcal{C}A_\mathrm{IPM}L^3_zI_0 g^{\rm pu}_{yzzy}g^{\rm pr}_{yzzy} \sin\Omega_\mathrm{IPM}t\nonumber\\
  &&\times A^\mathrm{lin}_\mathrm{eff}\sin\left(2\theta-\delta^\mathrm{lin}\right)
\label{eqn:S2S3a}
\end{eqnarray}

\noindent with
\begin{equation}
\delta^\mathrm{lin} =\arctan\left(\frac{\sqrt{3}\left(A_2-A_3\right)}{4A_1+A_2+A_3}\right)
\label{eqn:deltalin}
\end{equation}

\noindent and
\begin{equation}
A^\mathrm{lin}_\mathrm{eff} = \frac{4A_1+A_2+A_3}{4\cos\delta^\mathrm{lin}}.
\label{eqn:Aefflin}
\end{equation}

\begin{figure}[b]
	\includegraphics[width = \columnwidth]{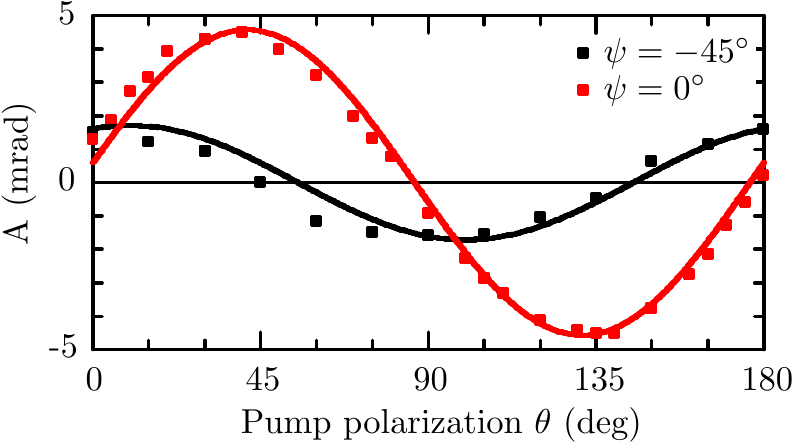}
	\caption{\label{fig:domaindistribution} Amplitude of the in-plane mode as a function of pump polarization angle $\theta$ for $\psi = -45^\circ$ and $0^\circ$. The dependence can be explained by contributions from different $S$-domain states.}
\end{figure}

Let us now analyze the distribution of the domains probed. For a single $S_1$-domain, the amplitude for $\psi = \pm45^\circ$ would be zero for all pump angles. However, the amplitude of the in-plane-mode as a function of pump polarization for different detection angles (Fig.~\ref{fig:domaindistribution}) immediately reveals the presence of a multi-$S$-domain composition of the sample. A fit of Eqs.~(\ref{eqn:S2S3}) yields
\begin{eqnarray}
A_1 = 0.651\pm 0.004\nonumber\\
A_2 = 0.064\pm 0.007\\
A_3 = 0.285\pm 0.012,\nonumber
\label{eqn:Areafractions}
\end{eqnarray}

In combination with Eqs.~(\ref{eqn:deltalin}) and (\ref{eqn:Aefflin}), we find
\begin{equation}
\delta^\mathrm{lin} = -7.4^\circ\pm0.7^\circ,
\end{equation}
\begin{equation}
A^\mathrm{lin}_\mathrm{eff} = 0.744\pm0.002.
\end{equation}

\noindent As anticipated, the combination of $S_1$-, $S_2$-, and $S_3$-domains leads to a phase shift $\delta^\mathrm{lin}$ in the polarization dependence.

A similar analysis of the $S$-domain composition can be applied for the excitation of the in-plane mode with \textit{circularly polarized light}, that is, via the IFE. In analogy to Eq.~(\ref{eqn:S2S3}), we obtain
\begin{eqnarray}
\Delta\eta^{\sigma^\pm}_\mathrm{IPM}
 &=& \mathcal{C}L^2_zI_0 k^{\rm pu}_{yzx}g^{\rm pr}_{yzzy}
\cos\Omega_\mathrm{IPM}t\nonumber\\
 &\times& [A_1\cos\left(2\psi\right)\nonumber\\
 &+&      A_2\cos\left(2\psi-120^\circ\right)\nonumber\\
 &+&      A_3\cos\left(2\psi-240^\circ\right)],
\label{eqn:S2S3b}
\end{eqnarray}

\noindent which can be expressed as
\begin{eqnarray}
\Delta\eta^{\sigma^\pm}_\mathrm{IPM}&=&\mathcal{C}L^2_zI_0 k^{\rm pu}_{yzx}g^{\rm pr}_{yzzy} \cos\Omega_\mathrm{IPM}t\nonumber\\
&&\times A^{\sigma^\pm}_\mathrm{eff}\cos\left(2\psi-\delta^{\sigma^\pm}\right)
\label{eqn:S2S3c}
\end{eqnarray}

\noindent with
\begin{equation}
\delta^\mathrm{\sigma^\pm} =\arctan\left(\frac{\sqrt{3}\left(A_2-A_3\right)}{2A_1-A_2-A_3}\right)
\label{eqn:deltaIFE}
\end{equation}

\noindent and
\begin{equation}
A^\mathrm{\sigma^\pm}_\mathrm{eff} = \frac{2A_1-A_2-A_3}{2\cos\delta^\mathrm{\sigma^\pm}}.
\label{eqn:AIFE}
\end{equation}

In revisiting Fig.~\ref{fig:circ}, the dominance of the in-plane mode for $\psi\,=\,0$ and its small amplitude of approximately \SI{0.05}{mrad} for $\psi\,=\,45^\circ$ are striking. They point to the pronounced prevalence of the $S_1$ domain state so that, even without an explicit fit of Eq.~(\ref{eqn:S2S3b}), we can conclude that $\delta^\mathrm{\sigma^\pm} \approx 0$ and $A^\mathrm{\sigma^\pm}_\mathrm{eff} \approx 1.0$ in the probed area.

A refinement of our analysis by taking $S$-domain distributions into account as described in this section enables, in the following, quantitative statements about the strength of the magneto-optical coupling constants in NiO to be made.

\subsection{Magneto-optical coupling constants}\label{sec:couplingconstants}

This section focuses on the quantitative analysis of the magneto-optical coupling tensors $k_{ijk}$ and $g_{ijkl}$.

During the analysis of the out-of-plane mode given in Section \ref{sec:lin}, the fitting parameters $X_1$ and $X_2$ were extracted, which are directly related to the magneto-optical coupling constants $g_{yyxz}$ and $g_{zzxz}$. The extracted values yield $g_{zzxz}/g_{yyxz} \approx -3.6$. This is a significant deviation from the isotropic case with $\bar{3}m$ symmetry, where the ratio would be $-1$.\cite{Cracknell69,Eremenko92} This is strong confirmation that, although the deviation from the \textit{crystallographic} point symmetry $\bar{3}m$ toward $2/m$ by magnetostriction from the $S$-domains is small, the magneto-optical properties of NiO have to be discussed in the framework of the \textit{magnetic} point symmetry $2/m$.

Furthermore, by comparing the oscillation amplitude of the in-plane mode in Figs.~ \ref{fig:linIPM} and \ref{fig:circ}, the magnon generation efficiency via IFE and ICME can be compared. The pump fluences were \SI{80}{mJ/cm^2} in both cases. In Fig.~\ref{fig:linIPM}, the magnon was excited by the ICME with a maximum oscillation amplitude $l^\mathrm{ICME}_y$ of approximately \SI{4.7}{mrad}. In contrast, for generation via the IFE (Fig.~\ref{fig:circ}(a)), the observed oscillation had an amplitude $l^\mathrm{IFE}_y$ of \SI{0.13}{mrad}. In both cases, the dynamics were probed via the contribution of $l_y$ to the Cotton-Mouton effect. The quantitative evaluation of the two excitation paths is hindered, however, by the multi-$S$-domain distribution. With the analysis of the previous section, we can now renormalize the measured amplitudes for single-$S$-domain samples.

From Eq.~(\ref{eqn:S2S3a}) and (\ref{eqn:S2S3c}), we see that the ratio of the spin precession amplitudes is determined by
\begin{equation}
\frac{l^\mathrm{ICME}_y}{l^\mathrm{IFE}_y} = A_\mathrm{IPM}\frac{A^\mathrm{lin}_\mathrm{eff}}{A^{\sigma^\pm}_\mathrm{eff}}\frac{L_zg^{\rm pu}_{yzzy}}{k^{\rm pu}_{yzx}}.
\label{eqn:ratio}
\end{equation}

\noindent The anisotropy factor $A_\mathrm{IPM}$ can be derived from the exchange field\cite{SatohPRL10} $H_E = 2\pi\cdot\SI{27.4}{THz}/\gamma$ and the angular frequency of the mode $\Omega_\mathrm{IPM} = 2\pi\cdot\SI{0.14}{THz}$ according to\cite{Kalashnikova08}
\begin{equation}
A_\mathrm{IPM} = \frac{2\gamma H_E}{\Omega_\mathrm{IPM}} \approx 400.
\end{equation}

\noindent The second factor in Eq.~(\ref{eqn:ratio}) is geometric and accounts for the distribution of $S$-domains within the probed area. With our previously determined values for $A^\mathrm{lin}_\mathrm{eff}$ and $A^{\sigma^\pm}_\mathrm{eff}$, we conclude that the ratio of the induced magnetizations is $L_zg^{\rm pu}_{yzzy}/k^{\rm pu}_{yzx} \approx 0.1$. Consequently, the ICME induces a magnetization, which is about an order of magnitude smaller than that of the IFE in NiO. Even though NiO is structurally different, this value is in line with the values obtained by Raman scattering in rutile structure antiferromagnets.\cite{Lockwood12} Nevertheless, in NiO, this is overcompensated by the pronounced magnetic anisotropy so that in total the amplitude ratio of the induced magnon oscillation on a single $S$-domain equals $A_\mathrm{IPM}L_zg_{yzzy}/k_{yzx} \approx 50$.

Moreover, we can consider the magnetic anisotropy energy, which applies to the in-plane mode:
\begin{equation}
\mathcal{H}_\mathrm{aniso} = \frac{a}{2}m^2_x + \frac{b}{2}l^2_y.
\end{equation}

\noindent This anisotropy leads to an elliptical spin motion. Consequently, $m_x = 0$, when $l_y$ is maximized and vice versa. Therefore, the ratio of the energies pumped into the magnetic system by the ICME and the IFE scales with the square of the ratio of the $l_y$-amplitudes, which is about $50^2$, or 2500.

As the IFE and ICME are described by antisymmetric and symmetric tensors $k_{ijk}$ and $g_{ijkl}$, respectively, we can now revisit the apparent contradictions in earlier Raman scattering experiments.\cite{Grimsditch98} There, it had been argued that the commonly accepted antisymmetric Raman scattering tensor is not sufficient to explain their results, but a symmetric tensor would. Moreover, they estimated that the symmetric contribution would be dominant. This is now confirmed, explained and quantified by our measurements.

\section{Conclusions}\label{sec:conclusions}

We performed time-resolved pump-probe measurements of two magnon modes in antiferromagnetic NiO. Measurements were performed on $T_0$-domains on the (111) surface of the sample. Thus, pump and probe pulses were propagating along the optic axis of the crystal, which avoids loss of the initial light polarization due to birefringence. This allowed us to study the dependence of the amplitude and phase of the induced magnon oscillations on pump polarization in detail. Comparing the measurements to an analytical model under consideration of the full magnetic $2/m$ point symmetry, we clarified the driving force of the individual magnon modes. Our model predicts clear selection rules for the dependence of the optical response on the probe conditions, which were verified in experiments.

The ICME constitutes the excitation mechanism for both the in-plane and the out-of-plane magnon modes by linearly polarized light. Its analysis even provides highly sensitive quantitative access to the distribution of the elusive $S$-domain sub-structure of the otherwise dominating $T$-domain distribution.

When circularly polarized pump pulses are used, the general behavior of the in-plane mode is qualitatively different from the out-of-plane mode. Such pulses propagating along the $x$-axis excite the out-of-plane mode via the ICME; the IFE becomes the driving mechanism of the in-plane mode. Comparison of the amplitudes of the magnon oscillations resulting from ICME and IFE revealed that the energy transfer into the magnetic system via the ICME is about three orders of magnitude more efficient than via the IFE. Whereas the magneto-optical coefficients parameterizing the ICME are about an order of magnitude smaller than those of the IFE, the dynamics induced by the ICME are significantly more pronounced due to the strong magnetic anisotropy. This resolves the long-standing question about the proclaimed dominance of the second-order ICME over the first-order IFE derived from Raman scattering experiments.

\begin{acknowledgments}
T. S. was supported by KAKENHI (Grant Nos. 15H05454 and 2610304) and JST-PRESTO and thanks ETH Zurich for hosting him on a guest Professorship. C. T. and M. F. acknowledge support from the SNSF project 200021/147080 and by FAST, a division of the SNSF NCCR MUST.
\end{acknowledgments}

\bibliography{PRB2016}

\end{document}